# Continuous-Wave Multiphoton Photoemission from Plasmonic Nanostars

*Murat Sivis[1], Nicolas Pazos-Perez[2], Renwen Yu[3], Ramon Alvarez Puebla[2,4], F. Javier García de Abajo[3,4], and Claus Ropers[1]*

[1]University of Göttingen, 4th Physical Institute - Solids and Nanostructures, Göttingen, Germany

[2]Department of Physical Chemistry and EMaS, Universitat Rovira i Virgili, Carrer de Marcel•lí Domingo s/n, 43007 Tarragona, Spain

[3]ICFO—Institut de Ciencies Fotoniques, The Barcelona Institute of Science and Technology, 08860 Castelldefels (Barcelona), Spain

[4]ICREA—Institució Catalana de Reserca i Estudis Avançats, Passeig Lluís Companys 23, 08010 Barcelona, Spain

ABSTRACT Highly nonlinear optical processes, such as multiphoton photoemission, require high intensities, typically achieved with ultrashort laser pulses and, hence, were first observed with the advent of picosecond laser technology. An alternative approach for reaching the required field intensities is offered by localized optical resonances such as plasmons. Here, we demonstrate localized multiphoton photoemission from plasmonic nanostructures under continuous-wave illumination. We use synthesized plasmonic gold nanostars, which exhibit sharp tips with structural features smaller than 5 nm, leading to near-field-intensity enhancements exceeding $10^3$. This large enhancement facilitates 3-photon photoemission driven by a simple continuous-wave laser diode. We characterize the intensity and polarization dependencies of the photoemission yield from both individual nanostars and ensembles. Numerical simulations of the plasmonic enhancement, the near-field distributions, and the photoemission intensities are in good agreement with experiment. Our results open a new avenue for the design of nanoscale electron sources.

Nanoscale confinement of optical fields in plasmonic structures is accompanied by significant intensity enhancements that increase the strength of both linear and nonlinear phenomena, such as single-molecule Raman-scattering[1,2] and fluorescence[3], second-, third-, and fifth- harmonic generation[4-8], nanoplasma extreme-ultraviolet light generation[8-11], and multiphoton photoemission[12-20]. The latter has attracted much attention in experiments involving surfaces[18-20], resonant antennas[17], or sharp metallic tips[12-16], facilitating the development of high-coherence tip-based laser-driven pulsed electron sources[21-28] for time-resolved electron microscopy and diffraction[29-31].

In general, nonlinear optical signals can be enhanced by confining a given incident average power in time and/or space. While temporal confinement is ubiquitous in the use of ultrashort laser pulses, additional spatial confinement is realized in optical nanostructures, defining the field of *ultrafast nano-optics*[32,33]. In particular, extensive theoretical and experimental work has shown a growing level of control over the near-field localization associated with resonant modes in optimized nanostructure geometries[34,35]. Exceedingly large field-enhancements in plasmonic nanostructures suggest the observation of highly nonlinear processes even under continuous-wave (CW) illumination conditions.

Here, we demonstrate nonlinear photoelectron emission from individual resonant gold nanostars under CW excitation at incident intensities below 1 $MWcm^{-2}$, using a 630-nm low-power laser diode. We characterize the CW multiphoton photoemission yield as a function of incident intensity and polarization, and further provide spatial scans to identify emission from individual nanostars. These findings are compared with photoemission measurements using 10 fs laser pulses at 800 nm central wavelength. Additionally, we present simulations of the electromagnetic near-field distributions and the resulting photoelectron yield that further support the nanoscale plasmonic origin of CW nonlinear photoemission at the single-particle level. Our results illustrate the potential of plasmonic field confinement in tailored resonant nanostructures to widely proliferate nonlinear nano-optics beyond ultrafast science.

The nanostars used in our experiments are grown by a seed-mediated approach[36,37] (see Methods for details) and exhibit multiple protuberances terminating in sharp tips, with radii as small as 4 nm (insets to Figure 1a,b). Despite the particle-to-particle variability in the detailed nanostar morphology, the controlled growth conditions used in the synthesis allow us to tune their plasmonic response close to the laser operation wavelengths of either 660-nm (CW) or 800 nm (fs-pulses with 190 nm full-width-at-half-maximum (FWHM) spectral bandwidth). Figure 1c shows the measured ensemble optical extinction spectra for both sets of nanostars deposited on glass slides (solid curves). Electromagnetic simulations of individual nanosstars (Fig. 1d) from each sample batch, with structural features sizes extracted from the TEM images in Figs. 1a and b, yield spectra (dashed curves) agreeing well with the central wavelength of measured response function. The simulated spectra are essentially dominated by one of the protruding tips of the particle, and therefore, notably narrowed compared to the experimental ensemble spectra. The calculated intensity enhancement for single 3D nanostars, as presented in Figure 1d, exceeds $10^3$ at tip regions a few nanometers in diameter. Figure 1d plots the magnitude squared of the optical field component which is locally perpendicular to the surface, as the surface-parallel component does not contribute significantly to photoemission due to low quantum efficiency[38].

In the photoemission experiments, nanostars (cf. Figures 1a and 1b) dispensed on a fused silica substrate with conductive indium-tin-oxide (ITO) coating are illuminated with focused CW or fs-pulsed laser radiation (see Methods for details), as depicted in Figure 1e. The focal-spot diameters (FWHM of intensity) are 3.5 μm × 1.1 μm (major × minor axis) and 5 μm for the CW and femtosecond-pulsed illumination, respectively, enabling the excitation of single nanostars for samples with a surface coverage of 0.1 particles/μm². Polarization and intensity control is realized with a broadband half-wave plate and a thin-film polarizer. The photoemission measurements are conducted in a high-vacuum chamber at background pressures of $10^{-7}$ mbar. Emitted photoelectrons are detected using a phosphor-screen microchannel-plate (MCP), imaged by a charge-coupled device camera for a moderate bias voltage (-10 to -30 V) applied to the sample, drawing emitted electrons towards the grounded detector front plate. Spatial photoemission maps are obtained by scanning the samples relative to the laser focus using a precision 3D translation stage.

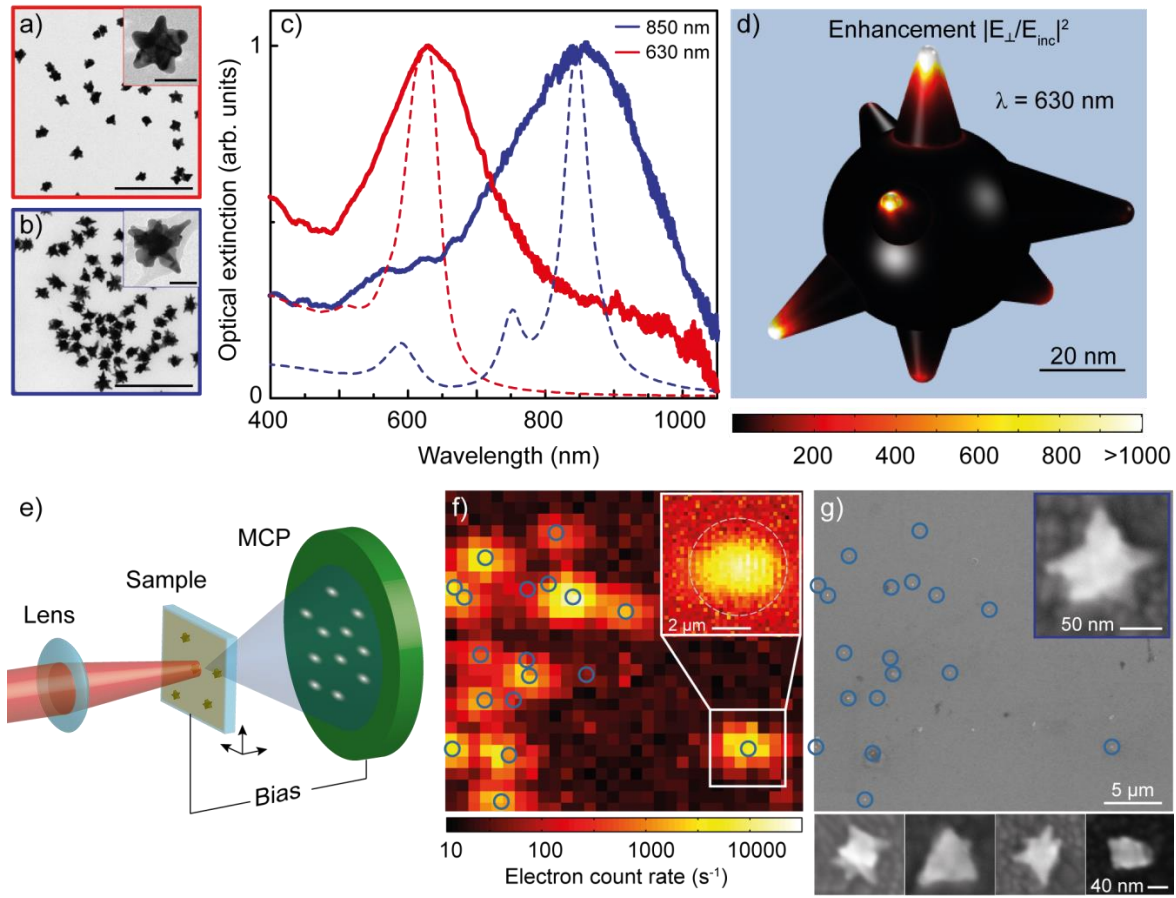

Figure 1 | **Optical field enhancement and femtosecond photoemission in resonant nanostars.** **a,b)** TEM images of nanostars on $Si_3N_4$ membranes (500 nm scale bars). Insets show close-ups of single stars (50 nm scale bars). **c)** Measured optical extinction spectra (solid lines) for ensembles of the two nanostar batches (see color-coded frames in a,b), compared to simulated spectra of individual nanostars from the respective batch (color-coded, dashed). **d)** Simulated intensity-enhancement (magnitude square of surface-normal electric field component) for a single nanostar with dimensions extracted from a. **e)** Schematic of the experimental setup: light is focused onto the sample (nanostars deposited on the glass substrate side facing towards the electron detector); electrons are detected using a microchannel plate (MCP) phosphor-screen assembly. **f)**

Photoemission map from a nanostar sample (850 nm resonance wavelength, surface density of 0.1 µm$^{-2}$). The inset shows a fine scan over a single star, with the dashed circle indicating the laser spot size (FWHM intensity). **g)** Scanning electron micrograph of the same region scanned in (f). The insets show close-ups of individual nanostars. The circles in f and g indicate the positions of nanostars leading to considerable photoemission.

Figures 1f and g show a photoemission map (recorded with femtosecond excitation) and a scanning electron micrograph of the scanned region on the nanostar sample (850 nm ± 100 nm resonance wavelength; see Figs. 1b, c), respectively. The photoemission hotspots can be clearly identified as positions of single or multiple nanostars, indicated by the blue circles. Only particles exhibiting nanometric feature sizes (confirmed by scanning electron microscopy, see insets to Fig. 1g) yield photoemission at an incident peak intensity of 100 MWcm$^{-2}$. A finer scan of an individual star (see upper-right insets to Figs. 1f,g) reveals that the emission profile is significantly narrowed compared to the intensity FWHM diameter of the focal spot of the incident beam (*cf.* bright emission region and dashed circle), illustrating the nonlinearity of the emission process.

The findings presented in Figure 1f are in line with previous experiments on multiphoton photoemission from plasmonic nanostructures using femtosecond excitation[12,14,17-20,39-41]. However, the observation of photoemission at peak-intensities in the MWcm$^{-2}$ range indicates the particularly high enhancement factors of these nanostars compared with nanotip structures or bow-tie antennas.

In the following, by employing the high field enhancement in the nanostars, we demonstrate multiphoton photoemission under CW illumination with sub-MW cm$^{-2}$ incident intensities. The nanostar sample that exhibits a resonance at 630 nm ± 75 nm wavelength (see Figs. 1a,c) is excited with the 660-nm CW-line from a laser diode with 60 mW maximum output power. Figure 2a presents the intensity scaling (double-logarithmic plot) of the photoemission from approximately 200 nanostars in a higher-density sample (open red circles, 70 stars/µm²) and a single nanostar of a sparse sample (open red triangles, 0.1 stars/µm²) positioned in the CW laser focus. The photoemission from the single nanostar (verified by polarization dependence, see below) is measured at the hotspot position indicated with the arrow in the spatial scan of Figure 2b. The hotspot extension and shape reflect the nonlinearity of the photoemission process and the oval shape of the focal spot, respectively. For comparison, we also plot in Figure 2a the light-intensity-dependent photoelectron yield from a single nanostar produced upon femtosecond-pulse excitation (filled black circles). The nonlinear scalings of the photoemission signals $\propto I^n$ with incident light intensity $I$ is indicated with dashed red ($n$=3) and black ($n$=5) lines for CW and femtosecond excitation, respectively.

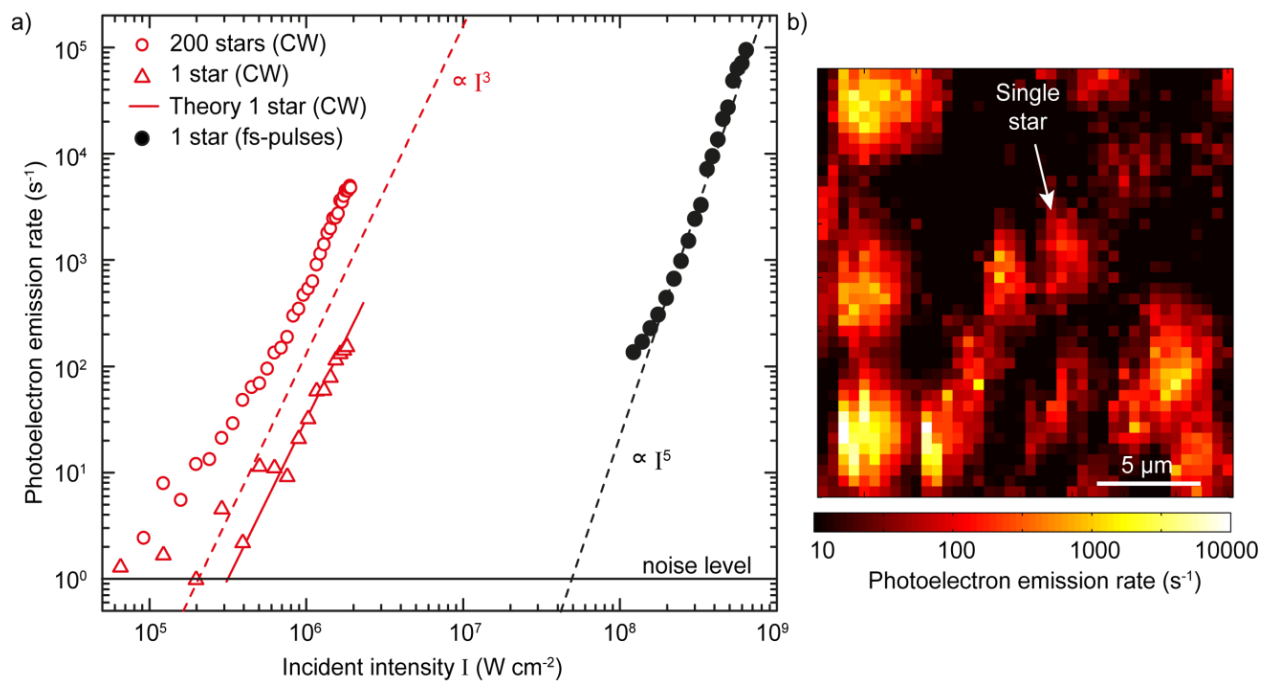

Figure 2 | **Continuous-wave multiphoton photoemission. a)** Light-intensity dependence of the photoemission yield from single and multiple nanostars. The solid line is the calculated 3-photon photoemission rate (see eq 1 in the Methods section) for a single nanostar under CW illumination, assuming an effective tip area of 5×5 nm$^2$ (*cf.* upper tip in Fig. 1d). The photoelectron emission rate for the experimental data is calculated from the electron count rate by assuming 10% detection efficiency. **b)** Photoemission map recorded with 45 mW CW excitation, corresponding to an incident intensity of 1.5 MWcm$^{-2}$.

In order to better understand the nonlinear photoemission process, we carry out perturbative simulations of the photoemission yield from a single nanostar under CW excitation (see solid red curve in Fig. 2a), based upon a description of conduction electrons as independent particles subject to a rectangular step potential to describe the surface barrier (see Methods for details). The results are in good quantitative agreement with the experimentally observed electron yield, justifying the employed perturbative treatment.

In both experiment and simulation, the far-field coupling to the resonant modes of individual nanostars strongly depends on the incident laser polarization. Figure 3 displays the polarization-dependent photoemission yields for three different stars, using either CW (solid red circles) or femtosecond-pulse (open black diamonds and blue circles) illumination. The measurements show a strong polarization-dependence of the photoemission yield for each individual nanostar, thus confirming a polarization-dependent mode coupling. We find that $(\cos \alpha)^{2n}$ fits of this polarization dependence (solid curves in Fig. 3) are consistent with the nonlinearity of the photoemission process for both CW ($n \approx 3$) and pulsed ($n \approx 4-5$) excitation. In the case of fs-excitation, some variation in nonlinearity for different nanostars likely stems from different resonance wavelengths overlapping the broad spectrum of the Titan:Sapphire laser (690-880 nm bandwidth) used, as well as local variations of the gold work function due to substrate effects, crystalline facets or the nanometric size of the features[42-44].

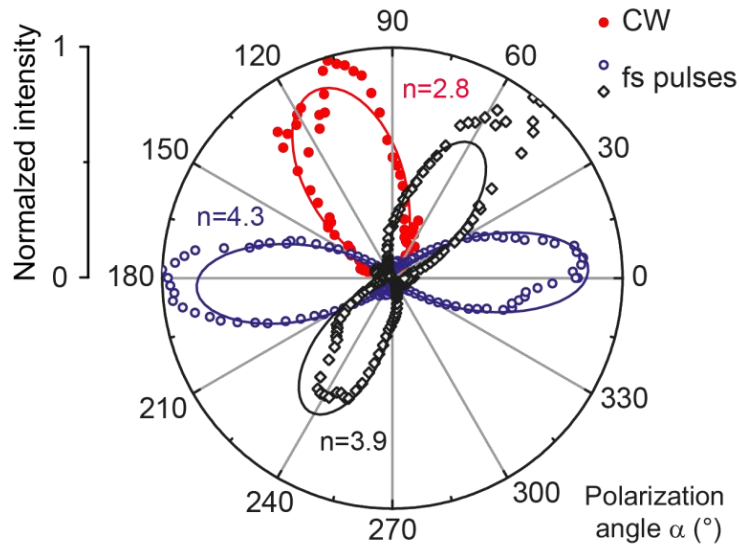

Figure 3 | **Polarization-dependence of photoemission from individual nanostars.** Measurements from three different nanostars, along with $(\cos \alpha)^{2n}$ fits (see values of *n* as color-coded labels) that show the nonlinear intensity dependence of the emission.

In conclusion, we demonstrated 3-photon photoemission from individual gold nanoparticles using low-power CW laser radiation at a wavelength of 660 nm (1.8 eV photon energy). This type of nonlinear processes requires large light intensities typically realized by employing ultrafast laser pulses. Instead, by harnessing a 1000-fold optical CW near-field intensity enhancement via localized plasmons at the tips of gold nanostars, we achieve a $>10^9$ fold total enhancement of the 3-photon electron yield, which agrees with calculations from a perturbative model. The findings suggest the use of very sharp tips (> 4 nm radii) as coherent electron sources in future nanoscale free-electron devices with potential applications in microscopy, spectroscopy, sensing, and signal processing.

METHODS

**Photoemission experiment**

Two sources are used to illuminate the samples with (i) few-femtosecond, nano-joule laser pulses having a central wavelength of 800 nm at 80 MHz repetition rate, and (ii) continuous-wave radiation at a wavelength of 660 nm from a low-budget (sub-100 €) laser diode.

**Gold nanostar synthesis**

*Materials and methods.* Poly(vinylpyrrolidone) (PVP, MW = 25 000) was purchased from Roth. Gold(III) chloride trihydrate (99.9%, $HAuCl_4 \cdot 3H_2O$), trisodium citrate dehydrated (≥99.5 %, $C_6H_5Na_3O_7 \cdot 2H_2O$), ethanol absolute (≥99.9%, EtOH), and indium tin oxide (ITO) coated glass slides (surface resistivity 8-12 Ω/sq) were obtained from Sigma-Aldrich. N,N-dimethylformamide (≥99%, DMF) was acquired from Fluka. All reactants were used without further purification. Milli-Q water (18 MΩ $cm^{-1}$) was used in all aqueous solutions, and all the glassware was cleaned with aqua regia before usage.

*Synthesis of spherical Au seeds nanoparticles.* Spherical Au nanoparticles of approximately 12 nm in diameter were produced by a modification of the well-known Turkevich method[45-49]. Briefly, Milli-Q water (500 mL) was heated to boil. After boiling had commenced, a solution of sodium citrate (11 mL, 0.1 M) was added to achieve a final citrate concentration of 2.2 mM. Boiling was continued for 10 min under vigorous stirring. After this time, 833.0 μL of an aqueous solution of $HAuCl_4$ (0.1 M) was added to the boiling solution and was left boiling under vigorous stirring during 30 min. A condenser was utilized to prevent the evaporation of the solvent. During this time, the color of the solution gradually changed from colorless to purple to finally become deep red. The resulting particles were coated with negatively charged citrate ions, and hence, were well suspended in H₂O. Next, after cooling, the particles were added drop by drop under stirring to a previously sonicated (30 min) aqueous solution of PVP (500 mL, 0.27 mM). To guarantee that adsorption was complete, the reaction mixture was stirred for 24 h at room temperature. Finally, the Au NPs were centrifuged (9000 rpm, 35 min) and redispersed (all in a total volume of 50 mL) in EtOH to achieve a final Au concentration of $16.2 \times 10^{-4}$ M.

*Synthesis of Au nanostars with $\lambda_{max}$ at 850 nm.* Au nanostars were prepared by a modification of a previously reported procedure[50] by dissolving 6.99 g of PVP in DMF (25 mL). After its complete dissolution, 10 mL extra of DMF were added, and the mixture was further sonicated for 30 min to assure homogeneity of the polymer in the solution. Followed by the addition of an aqueous solution of HAuCl₄ (77.7 µl, 0.1402 M) under rapid stirring at room temperature. Immediately after, 300 µL of the preformed dispersion of 12 nm, PVP-coated Au seeds in ethanol ([Au] = 16.2×10-4 M) was rapidly added. Within 15 min, the color of the solution changed from pink to blue, indicating the formation of Au nanostars. The solution was left under stirring overnight to assure the reduction of all reactants. DMF and excess of PVP was removed by several centrifugation steps, a first one at 7500 rpm for 40 min followed by two more at 7000 rpm for 10 min, in all steps the particles were resuspended in EtOH (35 mL). The obtained Au nanostars, exhibit a maximum absorbance peak at 850 nm.

*Synthesis of Au nanostars with $\lambda_{max}$ at 630 nm).* The previous synthesis of Au stars with $\lambda_{max}$ at 850 nm was repeated, and as before, after synthesis, the obtained Au nanostars were cleaned once by centrifugation (7500 rpm, 40 min) and redispersion in EtOH (35 mL). Next, two additions of HAuCl₄ (0.1402 M) each one of 15 µL were injected to the Au nanostar solution under vigorous stirring with 1 h time delay between them. After 4 h, the reaction was stopped by centrifugation

(2 × 7000 rpm, 10 min) and redispersion in EtOH (45 mL). The obtained Au nanostars, exhibit a maximum absorbance peak at 630 nm.

*ITO substrate preparation.* One-side coated ITO square glass slides (L × W × thickness = 25 mm × 25 mm × 1.1 mm) were bought and coated with 100 nm ITO on the non-coated glass side by sputtering at 150 W during 425 s.

*Au nanostars substrate deposition.* Both types of Au nanostars (850nm and 630 nm) were extra cleaned by four-fold centrifugation (7000 rpm, 10 min) and redispersion in EtOH. After that, two sets of solutions for each type of stars were prepared with final Au concentrations of $8\times10^{-5}$ M, $8\times10^{-4}$ M, and $8\times10^{-2}$ M. Next, 50 µL of each concentration were spin coated (1st ramp at 500 rpm for 10 s; 2nd ramp at 3000 rpm for 30 s with an acceleration rate for both ramps of 500 rpm/s) on ITO coated (on both sides) glass slides to achieve particle densities of approximately 0.1, 1.1, and 70 particles/µm$^2$ for the 630 nm Au stars and 0.07, 0.5, and 15 particles/µm$^2$ for the 850 nm Au Stars.

*Au nanostars deposition on SiN TEM grids.* Au Nanostars were deposited on a TEM SiN grid via spin coating (5 µL; 1st ramp at 500 rpm for 10 s; 2nd ramp at 3000 rpm for 30 s with an acceleration rate for both ramps of 500 rpm/s) from two different Au concentrations ($8\times10^{-5}$ M, $4\times10^{-4}$ M) to achieve particle densities of 0.18 and 0.6 particles/µm$^2$.

*Au nanostars deposition on glass slides for solid UV-VIS characterization.* Solutions of both types of Au Nanostars with concentrations of $5\times10^{-4}$ M were prepared and spin coated (50 µL, 500 rpm, 60 s) on microscope cover-slip glass slides to achieve a low particle density sufficient to avoid interparticle coupling while enabling UV-vis spectra to be recorded.

*Optical characterization.* UV-VIS spectroscopy was recorded with a PerkinElmer, Lambda 19. Size, shape, and topographical characterization of the nanoparticles and the substrates were performed with transmission and scanning electron microscopy (TEM, LEO 922 EFTEM operating at 200 kV and LEO 1530 FE-SEM, Zeiss).

**Theoretical methods**

*Electromagnetic simulations.* Extinction spectra and near-field distributions are calculated using a finite-difference method (COMSOL) to solve Maxwell's equations under external plane-wave illumination for characteristic nanostar morphologies (see Figure 1d). The dielectric function of gold is taken from optical data[51].

*Multiphoton photoemission.* An estimate of the photoemission yield is obtained by considering a flat surface exposed to a normal electric field with an amplitude given by the maximum intensity of the locally normal near-field resulting from the electromagnetic calculations for the nanostars. An effective hotspot area of 5×5 nm² is assumed (i.e., we multiply the electron emission current density by this area). The flat-surface approximation is justified by the small electron wavelength (~1 nm at the Fermi level of gold) compared with the nanostar tip rounding radius (~4 nm). We describe the gold flat-surface through a square-step potential (depth $V_0 = 16.3$ eV, work function $\Phi = 4.5$ eV). Available analytical solutions[41] for the initial, intermediate, and final electron states are used (see detailed explicit expressions for orthonormalized wave functions in Ref. 52), including their plane-wave dependence along the *x-y* directions parallel to the surface. As the parallel wave vector $k_\parallel$ is preserved during the emission process, we study transitions involving the perpendicular wave-function components, starting from an initial state $\varphi_{n=0}(z)$ (energy $\hbar\varepsilon_0$ relative to the valence band bottom), and with each of the three absorbed photons (frequency *ω*) producing a transition from $\varphi_{n-1}(z)$ (energy $\hbar\varepsilon_0 + (n-1)\hbar\omega$) to $\varphi_n(z)$ (with $n = 1 - 3$).

Approximating the electron-light Hamiltonian by $(-e\hbar/m\omega)[E_z(z)e^{-i\omega t} - E_z^*(z)e^{i\omega t}]\partial_z$ for an optical electric field $E_z(z)e^{-i\omega t} + E_z^*(z)e^{i\omega t}$ along the surface-normal direction $z$, the electron transitions under consideration can be described by iteration of the perturbative expression[53]

$$\varphi_n(z) = \frac{-e\hbar}{m\omega}\int dz'\, G_0^+(z,z',\varepsilon_0 + n\omega)E_z(z')\partial_{z'}\varphi_{n-1}(z'),$$

which gives the excited wave-function component produced from $\varphi_{n-1}$, whereas

$$G_0^+(z,z',\varepsilon_0 + n\omega) = \frac{m}{\hbar^2 k_n}\begin{cases} A_n e^{ik_n(z+z')} - ie^{ik_n|z-z'|}, & z,z' > 0 \\ B_n e^{ik_n z}e^{-ik_n' z'}, & z > 0, z' < 0 \\ B_n e^{-ik_n' z}e^{ik_n z'}, & z < 0, z' > 0 \\ \frac{k_n}{k_n'}\left[-A_n e^{-ik_n'(z+z')} - ie^{ik_n'|z-z'|}\right], & z,z' < 0 \end{cases}$$

is the forward electron Green function that satisfies the identity

$$\left[-\frac{\hbar^2 \partial_z^2}{2m} + V(z) - \hbar\varepsilon\right]G_0^+(z,z',\varepsilon) = -\delta(z-z')$$

for the square-step potential $V(z)$. Here, $A_n = -i(k_n - k_n')/(k_n + k_n')$, $B_n = -2ik_n/(k_n + k_n')$, $k_n' = \sqrt{2m(\varepsilon + n\omega)}/\hbar$ is the electron wave vector along $z$ inside the metal, and $k_n = \sqrt{2m(\varepsilon + n\omega - V_0/\hbar)}/\hbar$ is the normal wave vector in the vacuum side. Finally, integrating over initial states (i.e., over $\varepsilon$ and $k_\parallel$ in the range $0 < \varepsilon + \frac{\hbar k_\parallel^2}{2m} < E_F = V_0 - \Phi$), the photoelectron current per unit area with $n=3$ photons (under the assumption that $\hbar n\omega > \Phi$) reduces to

$$J = \frac{e^2}{4\pi m\hbar\,\omega^2}\int_{k_{min}'}^{k_{max}'}\frac{dk_0'}{k_n}\left(\frac{2mE_F}{\hbar^2} - k_0'^{\,2}\right)$$

$$\times \left| \int dz \left( A_n \theta(z) e^{ik_n z} + B_n \theta(-z) e^{-ik'_n z} \right) E_z(z) \, \partial_z \varphi_{n-1}(z) \right|^2, \quad (1)$$

where $k'_0$ is the initial electron wave vector inside gold, $k'_{\min} = \text{Re}\left\{ \sqrt{\frac{2mV_0}{\hbar^2} - \frac{2mn\omega}{\hbar}} \right\}$, and $k'_{\max} = \sqrt{2mE_F}/\hbar$.

**Corresponding Authors**

Murat Sivis, email: msivis@uni-goettingen.de

F. Javier García de Abajo, email: javier.garciadeabajo@nanophotonics.es

**Author Contributions**

M.S., F.J.G.A., R.A.P., and C.R. designed the study, based on an idea conceived by F.J.G.A.. N.P. synthesized the gold nanostars, prepared the samples and characterized the samples microscopically and spectrally. M.S. conducted the photoemission experiments and analyzed the data. R.Y carried out the theoretical simulations. F.J.G.A worked out the photoemission theory. M.S., C.R., and F.J.G.A. wrote the manuscript with contributions from all authors. All authors have given approval to the final version of the manuscript.

ACKNOWLEDGMENT This study was supported in part by the Deutsche Forschungsgemeinschaft (DFG) within DFG-SFB755 "Nanoscale Photonic imaging" (project C08), the Spanish Ministerio de Economia y Competitividad (MAT2014-59096-P, SEV2015-0522, CTQ2014-59808R and RYC-2015-19107), the European Research Council (Marie Curie Actions FP72014-623527), the Generalitat de Catalunya (2014-SGR-480), AGAUR (2014 SGR 1400), and Fundació Privada Cellex.

ABBREVIATIONS CW, continuous wave; FWHM, full-width-at-half-maximum

REFERENCES

1. Nie, S.; Emory, S. R. *Science* **1997,** 275, 1102-1106.

2. Rodríguez-Lorenzo, L.; Álvarez-Puebla, R. A.; Pastoriza-Santos, I.; Mazzucco, S.; Stéphan, O.; Kociak, M.; Liz-Marzán, L. M.; García de Abajo, F. J. *J. Am. Chem. Soc.* **2009,** 131, 4616-4618.

3. Kühn, S.; Håkanson, U.; Rogobete, L.; Sandoghdar, V. *Phys. Rev. Lett.* **2006,** 97, 017402.

4. Bouhelier, A.; Beversluis, M.; Hartschuh, A.; Novotny, L. *Phys. Rev. Lett.* **2003,** 90, 013903.

5. Neacsu, C. C.; Reider, G. A.; Raschke, M. B. *Phys. Rev. B* **2005,** 71, 201402.

6. Hanke, T.; Krauss, G.; Träutlein, D.; Wild, B.; Bratschitsch, R.; Leitenstorfer, A. *Phys. Rev. Lett.* **2009,** 103, 257404.

7. Lippitz, M.; van Dijk M. A.; Orrit, M. *Nano Lett.* **2005,** 5, 799-802.

8. Sivis, M.; Duwe, M.; Abel, B.; Ropers, C. *Nat. Phys.* **2013,** 9, 304-309.

9. Sivis, M.; Duwe, M.; Abel, B.; Ropers, C. *Nature* **2012,** 485, E1-E2.

10. Sivis, M.; Ropers, C. *Phys. Rev. Lett.* **2013,** 111, 085001.

11. Iwaszczuk, K.; Zalkovskij, M.; Strikwerda, A. C.; Jepsen, P. U. *Optica* **2015,** 2, 116-123.

12. Ropers, C.; Solli, D. R.; Schulz, C. P.; Lienau, C.; Elsaesser, T. *Phys. Rev. Lett.* **2007,** 98, 043907.

13. Schenk, M.; Krüger, M.; Hommelhoff, P. *Phys. Rev. Lett.* **2010,** 105, 257601.

14. Bormann, R.; Gulde, M.; Weismann, A.; Yalunin, S. V.; Ropers, C. *Phys. Rev. Lett.* **2010,** 105, 147601.

15. Kruger, M.; Schenk, M.; Hommelhoff, P. *Nature* **2011,** 475, 78-81.

16. Park, D. J.; Piglosiewicz, B.; Schmidt, S.; Kollmann, H.; Mascheck, M.; Lienau, C. *Phys. Rev. Lett.* **2012,** 109, 244803.

17. Dombi, P.; Hörl, A.; Rácz, P.; Márton, I.; Trügler, A.; Krenn, J. R.; Hohenester, U. *Nano Lett.* **2013,** 13, 674-678.

18. Teichmann, S. M.; Rácz, P.; Ciappina, M. F.; Pérez-Hernández, J. A.; Thai, A.; Fekete, J.; Elezzabi, A. Y.; Veisz, L.; Biegert, J.; Dombi, P. *Sci. Rep.* **2015,** 5, 7584.

19. Tsang, T.; Srinivasan-Rao, T.; Fischer, J. *Phys. Rev. B* **1991,** 43, 8870-8878.

20. Aeschlimann, M.; Schmuttenmaer, C. A.; Elsayed-Ali, H. E.; Miller, R. J. D.; Cao, J.; Gao, Y.; Mantell, D. A. *Journ. Chem. Phys.* **1995,** 102, 8606-8613.

21. Hommelhoff, P.; Kealhofer, C.; Kasevich, M. A. *Phys. Rev. Lett.* **2006,** 97, 247402.

22. Yanagisawa, H.; Hafner, C.; Doná, P.; Klöckner, M.; Leuenberger, D.; Greber, T.; Hengsberger, M.; Osterwalder, J. *Phys. Rev. Lett.* **2009,** 103, 257603.

23. Hoffrogge, J.; Paul Stein, J.; Krüger, M.; Förster, M.; Hammer, J.; Ehberger, D.; Baum, P.; Hommelhoff, P. *J. Appl. Phys.* **2014,** 115, 094506.

24. Gulde, M.; Schweda, S.; Storeck, G.; Maiti, M.; Yu, H. K.; Wodtke, A. M.; Schäfer, S.; Ropers, C. *Science* **2014,** 345, 200-204.

25. Vogelsang, J.; Robin, J.; Nagy, B. J.; Dombi, P.; Rosenkranz, D.; Schiek, M.; Groß, P.; Lienau, C. *Nano Lett.* **2015,** 15, 4685-4691.

26. Schröder, B.; Sivis, M.; Bormann, R.; Schäfer, S.; Ropers, C. *Appl. Phys. Lett.* **2015,** 107, 231105.

27. Müller, M.; Kravtsov, V.; Paarmann, A.; Raschke, M. B.; Ernstorfer, R. *ACS Photon.* **2016,** 3, 611-619.

28. Storeck, G.; Vogelgesang, S.; Sivis, M.; Schäfer, S.; Ropers, C. *Struc. Dyn.* **2017**, 4, 044024.

29. Yang, D. S.; Mohammed, O. F.; Zewail, A. H. *P. Nat. Acad. Sci.* **2010,** 107, 14993-14998

30. Müller, M.; Paarmann, A.; Ernstorfer, R. *Nat. Comm.* **2014,** *5*.

31. Feist, A.; Bach, N.; Rubiano da Silva, N.; Danz, T.; Möller, M.; Priebe, K. E.; Domröse, T.; Gatzmann, J. G.; Rost, S.; Schauss, J.; Strauch, S.; Bormann, R.; Sivis, M.; Schäfer, S.; Ropers, C. *Ultramicroscopy* **2017,** 176, 63-73.

32. Li, K.; Stockman, M. I.; Bergman, D. J. *Phys. Rev. Lett.* **2003,** 91, 227402.

33. Aeschlimann, M.; Bauer, M.; Bayer, D.; Brixner, T.; Garcia de Abajo, F. J.; Pfeiffer, W.; Rohmer, M.; Spindler, C.; Steeb, F. *Nature* **2007,** 446, 301-304.

34. Alvarez-Puebla, R.; Liz-Marzán, L. M.; García de Abajo, F. J. *J. Phys. Chem. Lett.* **2010,** 1, 2428-2434.

35. Schuller, J. A.; Barnard, E. S.; Cai, W.; Jun, Y. C.; White, J. S.; Brongersma, M. L. *Nat. Mater.* **2010,** 9, 193-204.

36. Rodríguez-Lorenzo, L.; Álvarez-Puebla, R. A.; Pastoriza-Santos, I.; Mazzucco, S.; Stéphan, O.; Kociak, M.; Liz-Marzán, L. M.; García de Abajo, F. J. *J. Am. Chem. Soc*. **2009**, 131, 4616-4618.

37. Morla-Folch, J.; Guerrini, L.; Pazos-Perez, N.; Arenal, R.; Alvarez-Puebla, R. A. *ACS Photon.* **2014,** 1, 1237-1244.

38. Lompre, L. A.; Thebault J. *Appl. Phys. Lett.* **1975,** 27, 110.

39. Fecher, G. H.; Schmidt, O.; Hwu, Y.; Schönhense, G. *J. Electron. Spectrosc.* **2002,** 126, 77-87.

40. Kubo, A.; Onda, K.; Petek, H.; Sun, Z.; Jung, Y. S.; Kim, H. K. *Nano Lett.* **2005,** 5, 1123-1127.

41. zu Heringdorf, F. J. M.; Chelaru, L. I.; Möllenbeck, S.; Thien, D.; Horn-von Hoegen, M. *Surf. Sci.* **2007,** 601, 4700-4705.

42. Chelvayohan, M.; Mee, C. H. B. *J. Phys. C* **1982,** 15, 2305.

43. Bröker, B.; Blum, R. P.; Frisch, J.; Vollmer, A.; Hofmann, O. T.; Rieger, R.; Müllen, M.; Rabe, J. P.; Zojer, E; Koch, N. *Appl. Phys. Lett.* **2008,** 93, 446.

44. Zhang, Y;, Pluchery, O.; Caillard, L.; Lamic-Humblot, A. F.; Casale, S.; Chabal, Y. J.; Salmeron, M. *Nano Lett.* **2014,** *15*(1), 51-55.

45. Turkevich, J.; Stevenson, P. C.; Hillier, J. *Discuss. Faraday Soc.* **1951,** 11, 55-75.

46. Enustun, B. V.; Turkevich, J. *J. Am. Chem. Soc.* **1963,** 85, 3317-3328.

47. Turkevich, J., *Gold Bull.* **1985**, 18, 86.

48. Turkevich, J., *Gold Bull.* **1985**, 18, 125.

49. Frens, G., *Nat. Phys. Sci.* **1973**, 20, 241.

50. Rodriguez Lorenzo, L.; J. *Am. Chem. Soc.* **2009**, 131, 4616–4618.

51. Johnson, P. B.; Christy, R. W. *Phys. Rev. B* **1972,** 6, 4370-4379.

52. Garcia de Abajo, F. J.; Echenique, P. M. *Nucl. Instrum. Methods Phys. Res. B* **1993,** 79, 15-20.

53. Messiah, A., *Quantum Mechanics*. North-Holland: New York, 1966.